\documentstyle [12pt]{article}
\def\references{\subsection*{REFERENCES}
\bgroup\parindent=0pt\parskip=\itemsep
\def\refpar{\par\hangindent=1.2em\hangafter=1}}
\def\endreferences{\refpar\egroup}

\def\@biblabel#1{\relax}
\def\@cite#1#2{#1\if@tempswa , #2\fi}
\def\reference{\relax\refpar}
\def\@citex[#1]#2{\if@filesw\immediate\write\@auxout{\string\citation{#2}}\fi
\def\@citea{}\@cite{\@for\@citeb:=#2\do
{\@citea\def\@citea{,\penalty\@m\ }\@ifundefined
{b@\@citeb}{\@warning
{Citation `\@citeb' on page \thepage \space undefined}}%
{\csname b@\@citeb\endcsname}}}{#1}}
\def\gsimeq
{\hbox{\raise0.5ex\hbox{$>\lower1.06ex\hbox{$\kern-1.07em{\sim}$}$}}}
\def\lsimeq
{\hbox{\raise0.5ex\hbox{$<\lower1.06ex\hbox{$\kern-1.07em{\sim}$}$}}}

\def\pn{\par\noindent}
\def\ss{\smallskip}

\begin{document}

\renewcommand{\thefootnote}{\fnsymbol{footnote}}

  \title{The variable ROSAT X-ray spectrum of the  \\
BL Lac 0716+714}

 \author {M. Cappi$^{1,2}$, A. Comastri$^2$, S. Molendi$^{2,3}$,\\
G.G.C. Palumbo$^{1,4}$, R. Della Ceca$^5$ \& T. Maccacaro$^6$}

\maketitle
\pn
   $^1$ Dipartimento di Astronomia, Universit\`a di Bologna,
             Via Zamboni 33, I-40126 Bologna, Italy
\pn
   $^2$ Max-Planck-Institut f\"ur extraterrestrische Physik,
         P.O. Box 1603, D-85740 Garching, Germany
\pn
   $^3$ European Southern Observatory, Karl Schwarzschild str. 2,
             D-85748 Garching, Germany
\pn
   $^4$ Istituto Te.S.R.E., CNR, Via Gobetti 101, I-40129, Bologna, Italy
\pn
   $^5$ Department of Physics and Astronomy, Johns Hopkins University,
    Baltimore, MD 21218, USA
\pn
   $^6$ Osservatorio Astronomico di Brera, Via Brera 28, I-20121 Milano, Italy
\pn

\vspace{2truecm}
{\it Monthly Notices of the Royal Astronomical Society: in press}
\pn

\vfill\eject

 \begin {abstract}

The BL Lac object 0716+714 is a powerful flat spectrum radio source detected
by CGRO-EGRET at $\gamma$-ray energies.
It has been observed for about 21000 s over two days by the X-ray telescope
onboard the ROSAT satellite and strong flux and spectral variations have been
detected. In the time scale  of
two days the source varied by a factor of $\sim 7$; during a sudden
($\sim 1000$ s) flux increase, the variation was 70\%.
Spectral analysis was performed first using the model independent hardness
ratio technique and
then the data were fitted using several standard spectral shapes.
Both analysis agree on the conclusion that 0716+714 exhibits spectral
variations related to source flux variability which may be
attributed to a ``soft excess" component most prominent when the source was
in a low state. The spectrum of 0716+714 steepened in the soft X-ray band
when the flux decreased while in the hard X-ray band remained constant.
A double power law model, representing steep synchrotron radiation at low
energies and flat inverse Compton radiation above $\sim$ 1 keV, best fits
both the low and high state.
\end{abstract}

\vfill\eject

 \section {INTRODUCTION}

\par
BL Lac objects are a class of Active Galactic Nuclei (AGNs) characterized by
weak or absent emission lines, a polarized optical and radio continuum and
dramatic variability at all wavelengths. With High Polarization Quasars (HPQs)
and Optically Violently Variable Quasars (OVVs) which display rather similar
continuum properties, they form a class of objects collectively known as
blazars. There is now a general agreement that for blazars the radio-to-optical
continuum is emitted by high energy electrons via synchrotron radiation
(Bregman 1990). This emission mechanism may occur in a relativistic jet
closely aligned with the observer's line of sight.
The radiation we observe, therefore, may be highly enhanced since it is beamed
along the jet's axis (K\"onigl 1981). The origin of the high energy continuum
in BL Lac objects, on the other hand, is not yet clear. It has been suggested
that two components due to synchrotron and Compton radiation mechanisms
may be present in the X-ray band. While the former should smoothly connect
to the low frequency portion of the spectrum which, because of radiative
losses, should steepen with increasing frequency, the latter should
cause flattening in the X-ray band with respect to lower frequencies
(see Maraschi, Maccacaro \& Ulrich (1989) for a summary of the field).
\par
Observations of X-ray selected BL Lac objects by the Einstein Observatory
(Schwartz \& Madejski 1989), EXOSAT and GINGA
have shown that their spectra are steep and convex, suggestive
of a synchrotron origin (Barr et al. 1989).
Classical radio selected BL Lacs, on the other hand, tend to have flatter,
and in some cases concave, X-ray spectra suggestive of a Compton radiation
component (Kii et al. 1991). Blazars in general have
been clearly detected at $\gamma$-ray energies by the EGRET instrument on the
Compton Gamma Ray Observatory (CGRO, Thompson et al. 1993).
These detections have been interpreted as the extension of the Compton
component to MeV energies and, in the case of MKN 421, to TeV energies (Punch
et al. 1993).
On average spectra of BL Lac objects observed by ROSAT (Fink et al. 1992)
have steeper slopes
than those measured by the Einstein Observatory at slightly higher
energies (Worral \& Wilkes 1990, Perlman et al. 1994).
\par
It should be noted that
X-ray observations of bright BL Lac objects with
high resolution instruments have sometimes revealed an absorption feature
at $E\sim 0.6\div0.7$ keV and $\Delta E\sim 100$ eV presumably due to O VIII
Ly$\alpha$ resonant absorption (Canizares \& Kruper 1984, Madejski et al.
1991).
The presence of such spectral feature, if not explicitely fitted,
could substantially modify the observed spectrum in a low resolution
detector like the ROSAT PSPC.
\par
Although it is known that BL Lac spectra do vary with time
(Giommi et al. 1990), it is not at all understood
whether the spectral variations are associated to the observed flux variations.
To test the validity of some model predictions, multifrequency
studies have become common practice. Simultaneous observations of the same
object at many different frequencies allow to explore different regions
and examine the physical processes at work.
The analysis of temporal and spectral variabilities
are, however, not always simple and the interpretation of the results is not
straightforward.
It must be emphasized, however, that observations as the ones discussed here
provide a powerful probe of the temporal and spectral behaviour of a well
defined physical region in the proximity of the central engine.
\par
In this paper the full analysis of a ROSAT observation of the BL Lac object
0716+714 is reported. This observation is of particular interest since
spectral and
temporal variability could be studied on rather long time scales, as the
ROSAT pointings were distributed over two days, as well as on short time
scales since the source experienced at least one outburst lasting about
1000 s during the observing period.
\par
The paper is organized as follows: general information on 0716+714 are
given in section 2 while the ROSAT observation is described in section 3.
Subsection 3.2 describes the temporal analysis and 3.3 the hardness ratio
analysis.
The spectral analysis is presented in subsection 3.4. A discussion of the
results is presented in section 4 and the conclusions are summarized in
section 5.
\par\noindent
Throughout the paper a Hubble constant $H_0=$ 50 km s$^{-1}$ Mpc$^{-1}$ and
a deceleration parameter $q_0=0$ are assumed.

\section {Data on BL Lac 0716+714}

The BL Lac object 0716+714 (RA: 07$^h$ 21$^m$ 53.4$^s$, Dec: 71$^d$ 20$^m$
36$^s$ at epoch 2000) is one of the BL Lacs best studied over the entire
electromagnetic spectrum.
It was discovered as the optical counterpart of one of the
1 Jy radio sources catalogued by K\"uhr et al. (1981) and belongs to the
complete sample of 1 Jy radio-selected BL Lacs (Stickel et al. 1991).
\par\noindent
In the radio band the overall spectrum is flat, with a spectral index
$\alpha_r = - 0.22$ between 11 and 6 cm ($S_{\nu} \propto \nu^{-\alpha}$),
and progressively steepens at higher energies (Impey \& Neugebauer 1988).
\par\noindent
In the optical band 0716+714 is the brightest object of the 1 Jy catalogue of
BL Lacs with $m_v = 13.2$ (Stickel et al. 1993) and is one of the five
objects in the sample which shows neither emission nor absorption features
in the spectrum even after long integration times.
A magnitude $m_v= 15.5$, corresponding to a decrease of $\Delta m_v=2.3$
between August 1979 and January 1980, has also been reported
(Biermann et al. 1981).
A lower limit on the redshift ($z > 0.3$)
has been derived from the non-detection of the host galaxy
(Wagner 1992).
\par\noindent
Intensive monitoring campaigns carried out simultaneously in the
radio and optical bands (Wagner 1992) show strong and correlated
variability occurring on time scales of less than one day.
\par\noindent
0716+714 is one of the five BL Lac objects detected by the EGRET
detector onboard CGRO at energies above 100 MeV.
Large flux and spectral variations have been detected at these
wavelengths (Kurfess 1994).
\par\noindent
The only previous X-ray measurement of 0716+714 in the X-ray band
has been carried out with the Einstein Observatory IPC in the $0.2-3.5$ keV
energy range (Biermann et al. 1981).
The source however was too weak to allow a detailed spectral and/or temporal
analysis.
\par

  \section {Data Analysis}

  \subsection{Data acquisition and reduction}

\par
The BL Lac object 0716+714 was observed on-axis with the PSPC detector
(Pfeffermann et al. 1986)
onboard the ROSAT Observatory (Tr\"umper 1983)
for a total of about 21000 s between the $8^{th}$ and $11^{th}$ of March 1991.
The analysis which follows is based on the data accumulated during this
observation obtained from the ROSAT public data archive.
The ROSAT telescope was pointed at the source 14 different times
but one pointing was so short that no photons were detected.
Source photons were collected from a circle of 2 arcmin radius centered on the
object position and background counts were collected from a $\sim$ 10 times
larger area in a ring surrounding the source. The total source photons
accumulated during the pointing was about 16300. The data were corrected for
the vignetting and deadtime of the telescope and the mean corrected count-rate
in the energy range $\sim 0.1-2.4$ keV was $0.802 \pm
0.006$ counts s$^{-1}$. Data preparation and analysis have been performed
using the JAN94 version of the EXSAS package (Zimmermann et al. 1993);
spectral analysis has been performed using version 8.33 of the XSPEC program
(Arnaud et al. 1991).

  \subsection{Temporal analysis}

\par
As a first step, all photons collected by ROSAT were binned
in 400 s time intervals as suggested by the wobble period of the telescope
and the resulting light curve is shown in Fig. 1, panel a.
The X-ray count-rate decreases with time from a high level of $\sim 1.8$ cts/s
to a low levelof $\sim 0.3$ cts/s two days later.
During the whole ROSAT observation, the ratio of maximum to minimum count rate
is $\sim 7$.
\par\noindent
In the second observation interval (5000-6000 s), an increase in the count rate
of about 70\% in less than 1000 s is present. This sudden burst of radiation,
hereafter called ``flare", has been previously noted by Witzel et al. (1993).
Panel b of Fig. 1 shows the increasing count rate during the flare;
the decay of this event was not monitored.
In a time span of 7000 s the source count rate decreased by a factor 2
(Fig. 1, panel c). As the count rate decreases with time the X-ray continue
to flicker but on longer time scales.

 \subsection{Hardness ratio analysis}

\par
Due to the high variability of this BL Lac, the first step in analyzing the
data has been to investigate variations in hardness ratio.
This provides a powerful tool for the detection of X-ray spectral
variability as the hardness ratio is model-independent.
\par\noindent
The hardness ratio $HR={(H-S)\over{(H+S)}}$ was computed using the EXSAS
standard
energy intervals: $S(\sim 0.1-0.4\ \rm{keV})=\rm{net\ counts\ between\
channels\ }
(11\div41)$ and $H(\sim 0.5-2.0\ \rm{keV})=\rm{net\ counts\ between\ channels\
} (52\div201)$.
\par\noindent
The full observation was subsequently devided in 4
different time periods and precisely the pre-flare (very
high and constant count rate), the flare (very high and increasing count rate),
the post-flare (high and decreasing count rate) and the low period
(low count rate).
Time selections, accumulated counts and mean count-rates for each period
are reported in Table 1.
\par\noindent

The $HR$ lightcurve (Fig. 2, panel a) clearly shows
that the first two intervals have harder $HR$ than the last
two. A $\chi^{2}$-test of the $HR$ light curve against constancy leads to a
$\chi^{2}_{red}\simeq 39.7$ for $3\ d.o.f.$ corresponding to a probability
$p\simeq 10^{-8}$.
\par\noindent
Given the high number of counts accumulated during the flare and post-flare
periods and the rapid flux variation seen,
each interval was further inspected for $HR$ variability; the results obtained
are consistent with $HR$ being constant within the statistical errors.
\par
{}From Fig. 2 panel a, two general states can be identified, each
with a constant hardness ratio:
a high (and hard) state at the beginning of the observation, including the
pre-flare and the flare periods (see Table 1) with $<HR>= -0.01\pm 0.02$, and
a low (and soft) state for the rest of the observation, including the
post-flare and the low period (see Table 1) with $<HR>=-0.12\pm 0.01$.
\par\noindent
In order to further investigate the spectral behaviour within the high and the
low state, a more detailed hardness ratio analysis has been performed.
For both states the data were binned in 4 energy ranges: the soft
band S1 (channels $11\div41$), the medium band S2 (channels $50\div89$),
the hard band H1 (channels $90\div139$) and the very hard band H2
(channels $140\div200$) (cf. Molendi \& Maccacaro 1994). Two new hardness
ratios defined as $HR_{soft}={(S2-S1)\over{(S2+S1)}}$,
$HR_{hard}={(H2-H1)\over{(H2+H1)}}$ have then been computed.
\par\noindent
The resulting hardness ratio light curves (Fig. 2, panels b and c)
show that the $HR_{soft}$ decreases with decreasing intensity
whereas the $HR_{hard}$ remains constant.
$\chi^{2}$-tests against constancy give probabilities of $\sim 2\cdot 10^{-5}$
and $\sim 0.66$ respectively.
In order to check whether the non-detected variation of $HR_{hard}$ were
related to lower statistics, the statistical errors associated to $HR_{hard}$
were assigned to $HR_{soft}$ and the resulting $HR_{soft}$
values tested against constancy; again a decrease of $HR_{soft}$ was detected
at a confidence level $> 3 \sigma$.
\par\noindent
The results from the hardness ratio analysis can be summarized as follow:
\par
1) The high state is harder than the low state at a confidence level $>>99\%$.
This is evidence for spectral variability associated with
amplitude variability, more specifically the source spectrum steepens
with decreasing intensity.
\par
2) The steepening is clearly observed below $\sim 0.9$ keV, while the
spectrum remains constant at harder energies, suggesting the presence of two
distinct spectral components over the broad energy band ($\sim 0.1-2.0$ keV).

 \subsection{Spectral analysis}

\par
To quantify the model independent but qualitative results obtained from
the $HR$ analysis, spectral analysis for the high
state and the low state (as defined in the previous section) was performed.
The spectra discussed in the following paragraphs were obtained from
the pulse height spectra binned
so that the signal to noise ratio would remain $\geq$ 10 in each channel;
the resulting degrees of freedom are, therefore, different for different
states. The ROSAT data reduction staff (ROSAT news letter
n.20, August 1993) recommended the use of the old matrix for AO-1 observations
such as the present one; to comply with these instructions the DRM06 detector
response matrix was used in the present analysis.

  \subsubsection{Single power law fits}

\par
A single power law model with low energy absorption has been
fitted to both states with free $N_H$.
The best-fit parameters are given in Table 2 and the
$\chi^2$ confidence contours in the parameter space $N_{H}-\Gamma$ are
shown in Fig. 3.
It is evident that the hydrogen column density derived
in the low state is not consistent at a confidence
level $>>99\%$ with the galactic value derived from radio
maps $N_{Hgal} = 3.95\cdot 10^{20} \rm{cm}^{-2}$ with an associated error of
the order of 10\%, see Dickey \& Lockman (1990).
On the contrary, the high state is reasonably well described by a single
power law with Galactic absorption.
\par
To test the possibility that the difference in $\chi^2_{red}$ for the low and
high state may be related to the difference in total counts accumulated,
only a part of the low state (the low period, with a total number
of counts $\sim$ 6900 which can be compared with the high state statistics)
was fitted with a power law model and Galactic absorption.
\par\noindent
A rather poor fit with $\chi^2_{red}\simeq2.2$ for 50 $d.o.f.$ was obtained
which corresponds to a probability $p\simeq 4\cdot10^{-6}$.
The resulting confidence contours are also shown in Fig. 3 and compared
with the results obtained for the low and high states.
A fit of the post-flare period yielded similar results
i.e. $\chi^2_{red}\simeq 1.6$ for 42 $d.o.f.$, corresponding to
$p\simeq 8\cdot10^{-3}$.
These results confirm that even with lower statistics, the low state is not
consistent with a single power law with Galactic absorption.
\par
This spectral analysis confirms the two previous results obtained with the
hardness ratio analysis:
\par
1) There is evidence for a spectral variation in 0716+714
related to the source variability.
\par
2) These variations can be attributed
to the presence of a second spectral component, a ``soft
excess", which is prominent only when the source is in the low state.

 \subsubsection{Two-components models}

\par
The pulse height spectrum of the low state was fitted
with the following two-component models: bremsstrahlung + power law,
double power law and power law + absorption trough.
$N_H$, when left free to vary,
turned out to have a value close to the Galactic value, therefore it
has been fixed at the Galactic value in all models.
\par
These three spectral models can be related to three different astrophysical
scenarios.
In the first model (thermal bremsstrahlung + power law), the soft thermal
component is assumed to be related to emission from the host galaxy while the
hard
power law component should be related to its nucleus. It could be argued that
if the galaxy hosting the BL Lac were an elliptical galaxy, a more realistic
model would be to fit a Raymond-Smith type of
spectrum. This fit was also attempted and the results turned out to be very
similar to the one obtained fitting a thermal bremsstrahlung spectrum.
However, as discussed later on, both models will have to be discarded on
the basis of luminosity considerations.
One should also point out that the soft portion of the spectrum can be
fitted by a black-body spectrum as well which, however, was disregarded
as of no physical meaning. This is to say that with the present data the
soft excess can set very weak constraints on models.
\par
In the second model (double power law), the steep power law dominating
at soft energies and the hard power law can be related respectively to
synchrotron and inverse Compton processes.
\par
In the third model (power law + absorption trough), the observed spectral
shape is explained as a single power law component undergoing absorption
by material present within the nucleus.
\par\noindent
With all the parameters free to vary, all three models yielded acceptable
fits to the source spectrum in the low state. The $\chi^{2}_{red}$ values are
respectively 0.78, 0.79 and 1.20 for 75 d.o.f. as reported in table 3
(``low state").
As an example the double power law model fit to the data is shown in Fig. 4.
Thus the present data do not allow to discriminate directly between the
3 proposed models.
\par
For the high state fit  as many parameters as possible were
fixed at the value found for the low state. The aim of this approach is
to select the model which could describe the source spectrum both in the low
and the high states varying the least number of parameters.
\par
The results obtained with the 3 different models are summarized in table
3 (``high state").
It should be  pointed out that the hardness ratio analysis was described
``going" from the high state to the low state while, for reasons of
consistency with the above mentioned approach, the results of the spectral
analysis are described ``going" from the low state to the high state.
\par
Moreover the following considerations seem in order:
\ss
\par
a) The bremsstrahlung + power law model apparently is the
model that best fits the data. In fact it describes both states by only
``shifting" up or down the power law normalization.
The spectral analysis shows that the pulse height spectrum can be
described in both low and high states with the sum of a constant plus
thermal emission and a variable power law which dominates at $E \ \gsimeq \ 1$
keV. The thermal bremsstrahlung has a characteristic temperature
$\sim$ 76 eV  and a constant flux $F_X(0.1-2.4) \simeq 7.3\cdot
10^{-13}\ \rm{ergs}\ \rm{cm}^{-2}\ \rm{s}^{-1}$. The power law
has a fixed steep spectrum with $\Gamma \simeq 2.7$ but varies
its intensity by a factor of about 3.
\par\noindent
In the context of this model, the spectral hardening with increasing flux
observed in the soft band (see section 3.4.1) can be understood in terms of
a larger contribution from the power law component, e.g. the hard component,
in the soft band with respect to its flux increase.
\ss
\par
b) The double power law model failed to describe both states when
only one normalization was varied ($\chi^{2}_{red}/d.o.f  \geq 7.4/33$).
This model requires the variation of at least 2 parameters in order to
explain the measured intensity and spectral variations.
Of the six possible choices with two free-parameters, only 3 yielded an
acceptable
fit for the high state and precisely the ones with $\Gamma_{hard}$ and
$A_{hard}$ free, $\Gamma_{soft}$ and $A_{soft}$ free or $A_{soft}$ and
$A_{hard}$ free (see table 3), where $\Gamma$ is the photon index
and $A$ is the power law normalization.
\par\noindent
The first solution is consistent with a variation in shape and
intensity of the hard component while the soft component is fixed.
This case requires an increase by a factor of $3\div4$ of the hard power law
intensity and a steepening of its photon index from $2.25$ to $2.72$.
\par\noindent
The second solution requires a spectral hardening with increasing flux
of the soft component while the hard component remains constant.
In this case, the soft component increases by a factor of
about 10 and shows a hardening ($\Gamma_{soft}$ going from 3.99 to
3.12). The soft component becomes the dominant component for
the whole broad energy band while the hard component remains constant.
\par\noindent
The third solution requires a flux variation of both components but
with this model, both components are constant in shape.
{}From the best fit values, we have obtained
${A_{soft}(high\ state)\over{A_{soft}(low\ state)}}\sim 2$ and
${A_{hard}(high\ state)\over{A_{hard}(low\ state)}}\sim 3$ .
These ratios show that the hard component increased more than the
soft component but both variations are of the same order of magnitude.
\par
All three possible double power law solutions described above
can explain the hardening in the soft band as measured with the
hardness ratio analysis (see section 3.3).
\par
c) The power law + absorption trough model failed to describe both the low
and the high states by only varying the covering factor,
giving $\chi^2_{red}/d.o.f. \simeq 44/33$ for the best-fit of the high state.
Therefore the observed intensity and spectral variations cannot
be explained simply varying the absorption trough.
With two free-parameters, namely the power law normalization and the
photon index, the model provides acceptable fits for both low and
high state. However, one should note that in this case the variations
are intrinsic and not related to the absorbing medium, and consequently
of little physical interest.
\ss
\par\noindent
The results of the spectral analysis of BL Lac 0716+714 can be summarized
as follows:
\par
1) There is a strong evidence for the presence of a second spectral
component in the energy band $\sim 0.1-2.4$ keV (section 3.4.1), at least
for the low state. This confirms the conclusions reached with the
hardness ratio analysis.
\par
2) The three different models used yield acceptable fits for the low state
spectrum and therefore, given the PSPC resolution, no preferential model
can be chosen.
\par
3) The attempt to describe both states with the same model allowes
to discard the power law + absorption trough model.
\par
In order to preserve a logical sequence of arguments the bremsstrahlung +
power law and the double power law models will be discussed in section
4.2 below.

  \section {Discussion}

  \subsection{Time Variability}

\par
For 0716+714 the shortest doubling time-scale detected in the soft X-ray band
is around 7000 s and occured during the post-flare period
(see Table 1 for definition).
During this time interval, the $0.1-2.4$ keV flux decreased by
$\sim 3.7\cdot10^{-11}$ ergs cm$^{-2}$ s$^{-1}$, indicating a
luminosity variation $\Delta L_{\rm x}\ \gsimeq \ 1.9\cdot 10^{46}$ ergs
s$^{-1}$
(assuming z $\gsimeq$ 0.3 and the double power law spectrum).
Assuming that the luminosity is produced by matter being transformed
into radiation with some efficiency $\eta$ ($\eta={{L\cdot t_{var}}\over
{M\cdot c^2}}$), the net result for a spherical source is
$$\Delta L\leq \eta {{m_p c^4}\over{\sigma_T}} t_{var}   \eqno (1)$$
\pn
where $\sigma_T$ is the Thomson cross section and $m_p$ the mass of the
proton (Fabian 1992).
\pn
The observed luminosity variation $\Delta L_{\rm x} \sim L_{\rm x}
\sim 1.9\cdot
10^{46}$ ergs s$^{-1}$ at $z = 0.3$ with $t_{var}\sim 7000/(1+z)$ s implies
$\eta > 1$, and therefore strongly suggests that the observed X-ray radiation
is beamed.
\par
Independent evidence of relativistic motion in this source is
derived from the observed superluminal velocity of radio knots
($\beta_{app} = 4.6$, Ghisellini et al. 1993). The same authors have
estimated a Doppler factor, in the framework of the
Synchrotron Self-Compton (SSC) model, of $\delta \geq 2.1$.
\par
A lower limit on the $\gamma$-ray luminosity has been computed from the
observed EGRET fluxes and spectral slopes
(Kurfess 1994).  In the $0.1-5$ GeV energy interval the observed
luminosity is in the range $L_{\gamma} \simeq 7.4 \cdot 10^{46} \div
1.2 \cdot 10^{47}$ ergs s$^{-1}$ (with z=0.3).
The fact that the $\gamma$-ray luminosity dominates the total radiated
power implies that 0716+714 must be transparent to photon-photon interaction.
The optical depth ($\tau_{\gamma \gamma}$) for this process is

$$ \tau_{\gamma \gamma}(x) = f(\alpha) {{l(1/x)} \over {4\pi}}  \eqno (2) $$

\par\noindent
where the compactness $l$ is defined as $l(x) = L(x) \sigma_T / R m_e c^3$,
$x$ is the photon energy ($x \equiv {{h \nu} \over {m_ec^2}}$),
$\alpha$ is the energy spectral index of the target X-ray photons in the
$\gamma$-ray emitting region and $f(\alpha)$ is a numerical factor taken
from Svensson (1987). In this process, a photon with energy
$x$ preferably interacts with photons of energy $1/x$.
\par
Assuming that $\gamma$-rays and X-rays are produced in the same region,
and imposing the condition of transparency for the photon-photon interaction
($\tau_{\gamma \gamma} < 1$), it is possible to compute the Doppler factor.
Estimating the intrinsic source size from $R = c t_{var} \delta$, one has
$\tau_{\gamma \gamma}(x) \propto \delta^{-4} L_{obs}(\delta^2/x)$
where $L_{obs}$ is the luminosity in the observer rest frame.
Assuming $L(\nu_0) \propto \nu_0^{-\alpha}$,
the limit on the source compactness (i.e. $\tau_{\gamma \gamma} < 1$) gives a
limit to the value of the Doppler factor:

$$ \delta> \left[ {f(\alpha) \sigma_T\over 4\pi c^2t_{var}}\,
       {\nu_0 L(\nu_0) \over h\nu_0} \right]^{1/(4+2\alpha)}   \eqno (3)  $$

\par\noindent
where $\alpha$, $\nu_0$ and $L(\nu_0)$ are the
energy index, the frequency and the monochromatic luminosity
of the X-ray target photons.
This formula, originally derived in Ghisellini (1993),
has been applied, in a somewhat different form, to 3C279 by Maraschi,
Ghisellini \& Celotti (1992).
{}From the observed doubling time scale and luminosity at
1 keV
($L_{\rm x}\simeq 1.3\cdot 10^{45} \rm{ergs}\ \rm{s}^{-1}\ \rm{keV}^{-1}$),
and with $\alpha = 1$ (leading to $f(1)\simeq 0.12$), the limit of
$\delta\ \gsimeq \ 3.2$
and $R\ \gsimeq \ 5.2 \cdot 10^{14}$ cm for the X and
$\gamma$-ray emitting region are obtained.
One should bear in mind that the above
values where obtained adopting a lower limit for the X-ray luminosity
estimate due to the conservative value ($z=0.3$) for the redshift.
\par
With the above values for the Doppler factor and the superluminal motion
$\beta_{app}\simeq 4.6$ an upper limit for the
Lorentz factor for the bulk motion $\Gamma \ \lsimeq \ 5.1$ and for the
angle to the line of sight $\phi \ \lsimeq \ 17$ degrees were obtained.

    \subsection {Spectral constraints}

\par
The hardness ratio and spectral analysis strongly suggest that
the soft X-ray spectrum of 0716+714 consists of two distinct
components. In section 3.4.2 we have compared the data to various
trial models.
The possibility that spectral and intensity
variations are related to variations of an absorption trough
has been shown to be inconsistent with the data.
\par
The thermal component + power law model can satisfactorily fit
the low and high state of the source by varying  only the
power law normalization. It would therefore appear that the
PSPC spectrum of 0716+714 can be explained as the sum
of a soft constant thermal component, related
to the host galaxy, plus a variable non-thermal component
associated to the BL Lac.
\par\noindent
The calculation of the lower limit
(assuming $z = 0.3$) of the luminosity for the thermal component
leads to a value of $L_{\rm x} (0.1-2.4$ keV) $\simeq 4 \cdot 10^{44}$ ergs
 s$^{-1}$.
This is more than two orders of magnitude larger than the highest
luminosity observed for elliptical galaxies (Fabbiano et al. 1992).
This model must, therefore, be abandoned as a plausible explanation
for the data.
\par
The only acceptable explanation for the X-ray spectrum is in terms
of a double power law model. In this context the steep power law
dominating at soft energies can be related to the synchrotron emission
while the flat power law emerging above $\sim 1$ keV is due to
inverse Compton emission.
The limited spectral resolution of the PSPC does not allow to
distinguish between different explanations for the observed spectral
variability. More specifically the low and the high state can be
satisfactorily fitted by varying: 1) the spectral index and normalization
of the soft component, 2) the spectral index and normalization of the
hard component, 3) the normalization of the soft and hard component.
All 3 possibilities are consistent with the SSC theory.
\par
The X to $\gamma$-ray spectral index
$\alpha_{{\rm x} \gamma} = - log (F_{\gamma}/ F_{\rm x})/ log (\nu_{\gamma}/
\nu_{\rm x})$
was computed, where $F_{\rm x}$ is the energy flux at 1 keV derived from the
data and $F_{\gamma}$
is the energy flux at 1 GeV derived from the EGRET observations (Kurfess 1994).
The obtained values are in the range $\alpha_{{\rm x} \gamma} \sim 0.8\div0.9$
depending on the considered X-ray or $\gamma$-ray states.
We note that both spectral indices for the hard X-ray power law component
($\alpha_{\rm x} = 1.25^{+0.40}_{-1.00}$)
and the $\gamma$-ray power law ($\alpha_{\gamma} = 0.8^{+0.2}_{-0.2}$)
are, within the errors, consistent with $\alpha_{{\rm x} \gamma}$.
Fig. 5 shows the multifrequency energy distribution of BL Lac 0716+714 with
the hard spectral behaviour emphasized.
It is tempting to speculate that the X to $\gamma$-ray spectrum of 0716+714
can be represented by a single power law of spectral index $\simeq$ 0.8$\div
$0.9, suggesting that both the ROSAT hard component
and the $\gamma$-ray emission could originate from the same mecanism, i.e.
inverse Compton scattering off relativistic electrons.

  \section {Conclusions}

\par
The main results of our paper can be summarized as follows:
\ss
\par\noindent
1) 0716+714 displays rapid variability at X-ray wavelengths,
more specifically a variation of a factor of about 7 in a time scale
of two days, a doubling time scale of $\sim 7000$ s
and a $\sim$ 70\% increase in $\sim$ 1000 s are observed.
\par\noindent
2) The strong $\gamma$-ray luminosity coupled with the rapid X-ray variability
has been used to derive a lower limit of $\delta > 3.2$ for the Doppler factor.
\par\noindent
3) The source clearly shows a steepening of the soft ($E \ \lsimeq \ 0.9$ keV)
X-ray spectrum with decreasing flux while the hard part of the spectrum
remains constant.
\par\noindent
4) The ``low state" spectrum of 0716+714 cannot be satisfactorily fitted
with a single absorbed power law model.
A double power law model, representing a steep synchrotron component
at low energies and a flat inverse Compton component dominating above $\sim$
1 keV, can describe both high and low state.
\par
The planned multifrequency campaign with frequency coverage also
at hard X-rays with the ASCA satellite (Fujimoto 1994)
will undoubtedly help to put stronger constraints on the spectral shape of
both components.

\section{Acknowledgements}

The authors wish to thank A. Celotti and G. Ghisellini
for many enlightening discussions on blazar emission models and for a critical
reading of the manuscript.
M.C. and A.C. thank Prof. Joachim Tr\"umper for the hospitality at MPE
and G. Zamorani for helpful discussions.
S.M. acknowledges financial support from MPE and ESO.
This work was made possible by the financial support from the European
Community, under the EEC contract No. ERB-CHRX-CT92-0033, MURST and ASI.

\clearpage
\samepage\onecolumn

\newpage\noindent

 {\large\bf FIGURE CAPTIONS}

\vskip 0.5 true cm
\pn
{\bf Fig. 1} ROSAT X-ray light curve of 0716+714 in the
0.1-2.4 keV energy range.
\pn
a) Full light curve showing the whole observation.
\pn
b) Blow-up of the ``flare".
\pn
c) Blow-up of a rapid flux decay.

\pn
{\bf Fig. 2} Hardness Ratios (see text for definitions) calculated along the
light curve of 0716+714.
\pn
a) The source is harder during the first two periods than during the last
two.
\pn
b) Hardness ratios calculated for the high and low states of the source
for the soft part of the spectrum.
\pn
c) The same as in b) but for the hard part of the spectrum.

\pn
{\bf Fig. 3} Absorption column density and photon index 68\%, 90\% and 99\%
confidence contour levels for both high and low states (solid lines).
Dashed lines represent the
confidence contours for the low period, to be compared with the high state
of equivalent statistic. The vertical line indicates the Galactic hydrogen
column density ($N_H = 3.95 \times 10^{20}$ cm$^{-2}$)

\pn
{\bf Fig. 4} ROSAT spectrum of 0716+714 in the low state (upper panel).
The data are fitted with a double power law model (Table 3). The residuals,
in the form of the ratio data/model, are shown in the lower panel.

\pn
{\bf Fig. 5} Multifrequency energy distribution for 0716+714.
Radio, microwave and infrared points are from K\"uhr et al. (1981), Stickel
et al. (1991) and Impey \& Neugebauer (1988) respectively. Optical data are
referred to an average value from Stickel et al. (1993) and Biermann et al.
(1981). The ultraviolet data are from IUE (Pian \& Treves 1993).
The X-ray data are from ROSAT (present work).
The soft and hard component slopes
are referred to the low state (see text). The $\gamma$-ray data are from
CGRO-EGRET (Kurfess 1994).

\end{document}